# Stock Index Futures Trading Impact on Spot Price Volatility.

# The CSI 300 studied with a TGARCH model

**Marcel AUSLOOS (*)**
School of Business,
University of Leicester, Brookfield,
Leicester, LE2 1RQ, United Kingdom
and
Department of Statistics and Econometrics,
Bucharest University of Economic Studies, Calea Dorobantilor 15-17,
Bucharest, 010552 Sector 1, Romania;

**Yining ZHANG (**)**
School of Business,
University of Leicester, Brookfield,
Leicester, LE2 1RQ, United Kingdom

**Gurjeet DHESI (***)**
School of Business
London South Bank University, 103 Borough Road
London, SE1 0AA, United Kingdom

**E-mail addresses :**
**(*)** ma683@le.ac.uk **; corresponding author; phone number: ++3243714340**
**(**)** zyn35302@hotmail.com
(***) dhesig@lsbu.ac.uk

**Abstract**

A TGARCH modeling is argued to be the optimal basis for investigating the impact of index futures trading on spot price variability. We discuss the CSI-300 index (China-Shanghai-Shenzhen-300-Stock Index) as a test case. The results prove that the introduction of CSI-300 index futures (CSI-300-IF) trading significantly reduces the volatility in the corresponding spot market. It is also found that there is a stationary equilibrium relationship between the CSI-300 spot and CSI-300-IF markets. A bidirectional Granger causality is also detected. "Finally", it is deduced that spot prices are predicted with greater accuracy over a 3 or 4 lag day time span.

# 1. INTRODUCTION

Supporters of futures markets are in favor of the view that futures trading contributes to enhancing the depth, "informativeness" and efficiency of the underlying stock market. Furthermore, they argue that risk aversion and price discovery functions in the futures markets also bring beneficial influences to stabilize the corresponding spot market (e.g. Stoll and Whaley, 1988; Harris, 1989). Other researchers state that futures trading increases spot market volatility due to a high degree of leverage and excessive speculative acts in futures markets (e.g. Furbush, 1989). Yet, there are reports about no significant volatility effect associated with the introduction of stock index futures. Thus, the results pertaining to volatility linkages between futures and spot markets are still controversial (Brailsford, Frino, Hodgson, and West, 1999; Afef and Olfa, 2009). In that respect, a recent thesis by Goetz (2019) is worth reading for a modern snapshot.

Possibly forecasting volatility is also an exciting ingredient of any research in such a field. For example, one can find studies on the dynamics of price volatility in the energy market, mainly the spot price of crude oil (Kristjanpoller and Minutolo, 2016; Quadry, Adeyemi, Jaiyeoba, and Alli, 2016) and the electricity market (Boland, Filar, Mohammadian, and Nazari, 2016), but also on « commodities » like gold (Kristjanpoller and Minutolo, 2015), or on the Atlantic salmon market (Asche, Misund, and Oglend, 2016).

Yet, most of the existing research concentrates on developed markets, mainly the USA and UK markets see a recent study by Antonakakis, Floros, and Kizys (2016); only a relatively small number of studies is associated with emerging markets: see Section 2 below. Our study focuses on analyzing the relationship between futures trading and spot price volatility in the Chinese stock market. "Recently", on April 8, 2005, the Shanghai and Shenzhen stock exchanges jointly issued the CSI 300 index. With the introduction of stock index futures on China markets, on April 16, 2010, empirical studies have doubted whether the onset of CSI 300 index futures brings a beneficial influence to stabilize the underlying spot market (Yang, Yang and Zhou, 2012; Wang and Xie, 2013; Zhang, 2015), one reason being the unique institutional trading structure in China (Miao et al., 2017; Huang and Ge, 2019); another cause proposed by Wu (2018) concerns the market maturity clauses.

Aiming to address such theoretical controversies on whether volatility spillover effects are present among spot and futures in the Chinese stock markets, and hopefully shining light on the controversies, precise questions are here below taken into account, from five practical points of view:
(RQ1) Is the spot market volatility influenced by stock index futures trading?
(RQ2) Does one see any "difference" in spot price fluctuations before and after the introduction of futures contracts?
(RQ3) Is there any possible predictability of prices, within a practical modeling?
Moreover, with further leaning toward theoretical grounds:
(RQ4) Is there (or not) a stationary equilibrium relationship between spot and futures markets?,
and finally,
(RQ5) Is there a marked causality relationship between the two markets, i.e. to what extent one market plays a more dominant role than the other?

In order to resolve these issues, as completely as possible, several kinds of time series modeling can be classically used: (i) Autoregressive Conditional Heteroscedastic (ARCH) models (Engle, 1982), (ii) GARCH model (Bollerslev, 1986), (iii) TGARCH model (Glosten, Jagannathan, and Runkle, 1993; Zakoian, 1994), (iv) EGARCH model (Nelson, 1991), (v) co-integration tests (Granger, 1981; Johansen, 1988, 1991), (vi) error correction models (Davidson et al., 1978), and (vii) Granger causality test (Granger, 1969). Such research methods are briefly recalled in Section 3 below, but the theoretical discussion ends in suggesting that the TGARCH way is the optimal, most complete, approach, to be used for the analysis.

Thus, Section 2 contains references to both theoretical or/and empirical literature, briefly discussing previous findings or open questions. Section 3 subsequently recalls research methods that are pertinent to *ad hoc* empirical analyses, but emphasizes the TGARCH model as the most pertinent. Section 4 explains the data collection and provides the descriptive statistics. Section 5 discusses the results of the empirical research. The co-integration tests, error correction models, and Granger causality test are found in Appendices.

Section 6 provides the conclusions. It is concluded that not only there is a stationary equilibrium relationship between the CSI 300 spot and CSI 300 IF markets, but also that the futures market affects the underlying spot market on a large basis, i.e. spot prices are predicted with greater accuracy by applying historical index futures prices. Thus, the findings allow us to conclude undoubtedly that index futures trading strengthens the efficiency, or stability, of the underlying spot market, in particular, this CSI 300 index, thereby giving much weight to the adequately proposed theoretical arguments.

## 2. LITERATURE REVIEW

Let us concentrate some attention upon previous empirical research associated with the relationship between prices in futures and spot markets. In brief, previous studies mainly discovered volatility linkages from three main angles: (1) the impact of stock index futures on spot market volatility; (2) the changes in stock market volatility before and after the introduction of index futures; (3) the comparison on volatility levels between stock index futures markets and spot markets. These findings are emphasized below.

Moreover, according to different conclusions in terms of the volatility linkages between futures trading and spot market, ideas and findings can be seen to fall into three classes: (1) increasing spot market volatility; (2) no change of price volatility in the stock market; (3) decreasing spot market volatility. Let us briefly focus on them.

### *GNMA and Stock index futures*

To maintain a proper perspective, recall that earlier works on financial futures mainly investigate the relationship between the Government National Mortgage Association (GNMA) futures trading and the price instability of the underlying cash market.

Froewiss (1978), through a regression analysis, found that the introduction of GNMA futures does not cause (weekly) an increase in spot price volatility. Simpson and Ireland (1982) reached a similar conclusion for daily and weekly price fluctuations.

In contrast, other researchers state that the introduction of the GNMA futures is rather useful in explaining the GNMA volatility in the cash market (Figlewski, 1981; Corgel and Gay, 1984; Box and Tiao, 1975；Doukas and Rahman, 1986) claiming that the introduction of the futures market exerts a long-run stabilizing impact on spot market volatility.

The impact of futures trading on GNMA cash market volatility was shown to be neutral (Bhattacharya, Ramjee and Ramjee, 1986).

*Stock index futures*

The influence of stock index futures on underlying stock market volatility can be found in "recent work". For example,

*(1) Increasing Volatility*

Furbush (1989) adopted a regression analysis to examine the relationship between types of program trading and S&P 500 stock index and futures prices over five-minute intervals for October 14-20, 1987. His research displays that program trading in the index futures market worsens the corresponding cash market volatility. The substantial increase of portfolio insurance in the futures market causes a great volume of program trading. Meanwhile, excessive program trading activities lead to violent fluctuations of stock prices during the October 19, 1987 stock market crash. Schwert (1989) examined the S&P 500 monthly data from 1857 to 1989. His results suggest that there is an increase in stock volatility from October 1987 to October 1989, while there is no significant change in price volatility before this period. Apparently, the destabilization of stock prices from October 1987 to October 1989 is mainly caused by the 1987 stock market crash. Thus, it seems that the inception of futures trading destabilizes the underlying spot market, by increasing its volatility, even after a crash.

A concomitant study by Cheung and Ng (1990) showed a negative volatility spillover effect between stock index futures and spot markets via various GARCH models when analyzing the price volatility from the 15-minute intraday observations of S&P 500. Furthermore, Cheung and Ng (1990) interestingly observed that the price volatility in the futures market has a 15 minutes lead over the stock market volatility. Such an intraday lead-lag relationship between index futures and spot markets has also been detected by Chan (1992) and by Lockwood and Linn (1990).

An increase in stock market volatility owing to the inception of stock index futures, among non-U.S. markets was found in the Japanese market, - through decomposing the Nikkei 225 (i.e. a Tokyo Stock Exchange stock market index) spot portfolio volatility into the cross-sectional dispersion and the average volatility of securities in the portfolios by Chang, Cheng, and Pinegar (1999). More recently, Yang et al. (2012) investigated price discovery and volatility transmission between index futures and spot markets employing co-integration analysis and an asymmetric Error Correction Model (ECM) with GARCH error on intraday high-frequency data for the period April 16, 2010 to July 30, 2010. Studies by Guo et al. (2013), Xie and Mo (2014), Nishimura and Sun (2015), and Zhou and Wu (2016) also indicate an increased volatility following the introduction of CSI 300 index futures. It will be pointed out that this is contrary to our findings.

*(2) Decreasing volatility*

On the stabilization side, i.e. decreasing market volatility, a few studies concern the U.S. stock market. Santoni (1987) discussed the changes in the S&P 500 daily and weekly price indices, before and after futures indices were introduced. His results show a negative relationship between trading activities in stock index futures and spot price variability. Thus, he concluded that an increase in futures volume leads to a decline in the destabilization of the spot market. Edwards (1988a, 1988b) assesses the argument that index futures trading destabilizes the cash market from measured of the variance of daily returns on the S&P 500 index for the period October 1, 1982 to December 31, 1986, thus, in line with Santoni (1987).

Maberly, Allen, and Gilbert (1989) work concerns the change in spot market volatility before listing stock index futures (from 01/01/1963 to 20/04/1982) and after launching stock index futures (from 21/04/1982 to 16/10/1987). Notice that, for us, Maberly et al. employ the most reasonable analysis approaches to examine whether there exists an increase in spot price volatility after the onset of S&P 500 index futures. At that time, Harris (1989) reported to have examined whether the presence of derivatives trading has a destabilizing effect on stock market volatility in adopting a cross-sectional analysis of covariance regression model between 1975 and 1987. A conclusion from Harris' paper implies that stock index futures increasing spot market volatility can be "only occasional". Harris' article presents the fundamental argument that trades in index futures markets make spot markets more efficient because prices are adjusting quickly to new information.

For non-U.S. markets, Robinson (1994) presented an analysis of daily price volatility on the London International Stock Exchange for the period 1980 to 1993 through ARCH-M (ARCH in mean) models. He concluded that the introduction of FTSE 100 index futures contracts significantly reduces spot market volatility, - by about 17 percent. Robinson's research indicates that there is no strong evidence to support the asymmetric responses of spot market volatility due to good or bad news. In other words, the introduction of stock index futures market is beneficial to promote the market stability. The hypothesis that futures trading has a significant effect in lowering the level of the underlying stock market volatility is also supported by Bologna and Cavallo (2002) for the Italian Stock exchange, by Pilar and Pafael (2002) for the Spain stock market, by Floros and Vougas (2006) for the Greek stock market, and by Kasman and Kasman (2008) for the Turkey stock market.

As for the impact of index futures trading in the Chinese stock market, Chen et al. (2013) analyzed the changes in stock market volatility before and after issuing index futures contracts (during 04/01/2002-30/06/2011) utilizing a recently advanced panel data evaluation approach on the daily returns of CSI 300 index. The authors similarly found that the establishment of CSI 300 index futures market significantly reduces the volatility, whence (positively) influences the spot price stability. Fan and Du (2017) found a "positive" spillover effect between CSI 500 index futures market and the spot market from high frequency data analysis (for 2015). Similar findings are in He and Jing (2017).

*(3) No influence or mixed results*

Other authors argue that there is no relationship between stock index futures trading and spot market volatility. Some studies concern the issue of the volatility effect in the U.S. (Aggarwal, 1988; Becketti and Roberts, 1990; Hodgson and Nicholls, 1991; Darrat and Rahman, 1995).

Some other evidence to support the hypothesis that there is no relationship between the onset of stock index futures and the volatility of the index returns can be also found for non-U.S. markets: almost no significant impact was documented for the Australia, Brazil, Hungary and Mexico market (Jochum and Kodres, 1998; Dennis and Sim, 1999). The latter paper uses an asymmetric exponential ARCH model for estimating spot price variability, together with Granger causality (1969) and Geweke feedback measures (1982, 1984, 1992). For the British stock market, Board el al. (2001) examined the controversy surrounding the relation between FTSE 100 stock index futures and the corresponding spot market via employing the GARCH and variance-volume models. The results indicate that neither futures trading nor the increase of futures volume in the futures market has any effect on changes in the volatility of FTSE 100 index returns.

Studying contract trading on the Karachi Stock Exchange and investigating the changes in the return volatility of the underlying stocks (using an augmented GJR-GARCH model as well the more traditional measures of return volatility) Khan, Shah and Abbas (2011) found "mixed results".

"Finally", Chen and Zhang (2015) studied the impact of CSI 300 index futures on the underlying spot market volatility through establishing a VAR model; from such an approach, their results show that index futures trading did not significantly affect the volatility of spot market. Chen and Zhang (2015) also indicated that the CSI 300 index futures price does not "Granger cause" the corresponding spot price. We will show that our findings are inconsistent with the conclusions of Chen and Zhang (2015).

Let us also stress that authors can find different results on the same data but with different methodologies. For example, Hu (2016) Based on the TARCH model, finds that the introduction of CSI 300 stock index futures reduces the asymmetric volatility of the stock market. However, based on the GARCH model, he finds that the price volatility of CSI 300 stock index futures has no significant effect on the volatility of the stock market.

*(4) Moreover distinguishing developing and developed markets or countries*

Besides, considering whether the spot price volatility is impacted by the introduction of an index futures or not, through a discussion of the methodology, one might filter the literature reports trough another lens, i.e. distinguishing emerging from developed markets. In fact, such an index futures in developed markets has usually been imagined some « long time » ago; this is not the case for most emerging markets. Among the most recent reports, one may find those pertaining to specific («country») markets, or reports comparing several countries.

In the former set, one finds, studies about Taiwan (Chiang and Wang, 2002), Korea (Ryoo and Smith, 2004), South Africa (Floros, 2009), Pakistan (Khan, Shah, and Abbas, 2011), India (Karthikeyan and Karthika, 2016), Thailand (Bamrungsap, 2018), Malaysia (Taunson et al., 2018) to which one can add Greece (Spyrou, 2005; Floros and Vougas, 2006; 2007) and Turkey (Kasman and Kasman, 2008).

In the latter set, one may refer to Yarovaya, Brzeszczyński, and Lau (2016), for 10 developed and 11 emerging markets, Alan, Karagozoglu, and Korkmaz (2016), comparing Turkish, Korean and Taiwanese markets, Aloui et al. (2018), for 6 emerging and 5 developed markets, Tarique and Malik (2018) on the BRICS economies, and Kutan et al. (2018), in 7 emerging countries, i.e., Turkey, Poland and BRICS,

In particular, Yarovaya et al. (2016) findings demonstrate that markets are more susceptible to domestic and region-specific volatility shocks than to inter-regional contagion. Notice the (for these topics) unusual method by Aloui et al. (2018) investigating the information transmission across stock indices and stock index futures through the wavelet technique.

In some way, this short review of much literature here above, pointing to findings, indicates that several questions remain open not only due to theoretical arguments but also due to the methodology.

## 3. METHODOLOGY

This short review of much literature here above indicates that several questions remain open not only due to theoretical arguments but also due to the methodology. We argue that a non-negligible missing ingredient concerns the asymmetric response of the prices to the news. Thus, we are led to adopt the TGARCH (Glosten et al., 1993; Zakoian, 1994) models. For completeness, it has to be coupled to co-; tests (Granger, 1981; Johansen, 1988, 1991), Error correction model (Davidson et al., 1978; Maddala, 2001), and the Granger causality test (Granger, 1969).

### 3.1 THE TGARCH MODEL

The autoregressive conditional heteroscedastic (ARCH) model basic concepts stand on "mean zero, serially uncorrelated processes with non-constant variances conditional on the past, but constant unconditional variances" (Engle, 1982, 987). Bollerslev (1986) removed the constraint on the conditional volatility through the "generalized ARCH" (GARCH) model. The conditional variance equation of the ARCH model is modified for a GARCH(p, q) process which reads then

$$Y_t = \beta' X_t + \varepsilon_t, \qquad \varepsilon_t | \psi_t \sim N(0, \sigma_t^2), \qquad (3.1.1)$$

$$\sigma_t^2 = \alpha_0 + \sum_{i=1}^{p} \alpha_i \varepsilon_{t-i}^2 + \sum_{j=1}^{q} \beta_j \sigma_{t-j}^2, \qquad (3.1.2)$$

with $\alpha_0 > 0$, $\alpha_i \geq 0$ $(i = 1, \ldots, p)$ and $\beta_j \geq 0$ $(j = 1, \ldots, q)$ to guarantee that the conditional variance is positive (Tsay, 2010; Asteriou and Hall, 2016); $\varepsilon_t^2$ denotes the square of residuals and $q$ the length of ARCH lags. The distribution of disturbances ($\varepsilon_t$) is assumed to derive from an information set at time *t*, denoted by $\psi_t$.

However, financial asset prices may respond asymmetrically to good and bad news in practice (Asteriou and Hall, 2016; Dhesi and Ausloos, 2016). Thus, Glosten et al. (1993) and Zakoian (1994) developed the "threshold GARCH" (TGARCH) model for such cases. In order to capture the asymmetric effect for either positive or negative news, one adds a multiplicative dummy variable into the conditional variance equation for investigating whether there is a statistically significant difference in the volatility $\sigma_t$ of the series (Zakoian, 1994). This specific form allows different reactions of the volatility to different signs of the lagged errors. The specifications for TGARCH (1,1) and more generally TGARCH (p, q) are given by

$$\sigma_t^2 = \alpha_0 + \alpha_1 \varepsilon_{t-1}^2 + \gamma_1 N_{t-1} \varepsilon_{t-1}^2 + \beta_1 \sigma_{t-1}^2, \qquad (3.1.3)$$

$$\sigma_t^2 = \alpha_0 + \sum_{i=1}^{p}(\alpha_i + \gamma_i N_{t-i})\varepsilon_{t-i}^2 + \sum_{j=1}^{q}\beta_j \sigma_{t-j}^2, \tag{3.1.4}$$

respectively. In equations (3.1.3)-(3.1.4), $N_{t-i}$ takes the value of 0 for $\varepsilon_{t-i} \geq 0$, and 1 for $\varepsilon_{t-i} < 0$ (Zakoian, 1994), such that the news shock is represented by a Heaviside step function at $\varepsilon_{t-i} = 0$, thus whatever the time *t-i*. It is assumed that $\alpha_i$, $\gamma_i$ $(i = 1, ..., p)$, and $\beta_j$ $(j = 1, ..., q)$ are nonnegative parameters. When $\varepsilon_{t-i}$ is positive, the impact of $\varepsilon_{t-i}^2$ on $\sigma_t^2$ (i.e. the conditional variance) is $\alpha_i \varepsilon_{t-i}^2$; when $\varepsilon_{t-i}$ is negative, there is a larger impact $(\alpha_i + \gamma_i)\varepsilon_{t-i}^2$ with $\gamma_i > 0$ (Tsay, 2010). In other words, the impacts of $\varepsilon_{t-i}^2$ on $\sigma_t^2$ is determined by whether $\varepsilon_{t-i}$ is above or below the threshold value of zero. According to equation (3.1.3), A positive shock has an effect of $\alpha_1$, whereas a negative shock has an effect of $\alpha_1 + \gamma_1$. If $\gamma_1 > 0$, there is asymmetry: bad news have a larger impact on volatility than good news (Zakoian, 1994). One might argue about such an assumption, and the threshold value restriction, but this is outside our present aim.

### 3.2. CO-INTEGRATION AND ERROR-CORRECTION MODEL

In addition to estimate the impact of CSI 300 stock index futures trading on spot market volatility, we have investigated whether there exists a long-run equilibrium relationship between spot and futures prices. Our research adopts co-integration tests for checking the presence of a long-run equilibrium relationship. It is known that prior to checking for co-integration of two time series, it is essential to examine whether both are stationary. The Augmented Dickey and Fuller (ADF) test, (Dickey and Fuller, 1979), is so used.

If the results of ADF tests indicate that both price series are non-stationary processes, co-integration tests can be carried out adopting the two-step procedure of Engle and Granger (1987) and the Johansen co-integration test (1991). The co-integration tests (Mandala (2001, pp. 556-560) are based on, XXXXX

$$Y_t = c + \beta' X_t + \varepsilon_t, \tag{3.2.1}$$

where $Y_t$ and $X_t$ will denote the CSI 300 stock index futures and the underlying spot price series, respectively, here below; $\varepsilon_t$ represents the disturbances.

Subsequently, we may apply the error-correlation model (ECM) proposed by Davidson et al. (1978). This model is employed for checking the short-run dynamics of the relationship between the two variables, spot and index futures prices. In brief, the ECM mechanism allows for correcting the disequilibrium of the previous period and for resolving the issue of spurious regression via eliminating stochastic trends in the variables (Tsay, 2010).

### 3.3. GRANGER CAUSALITY TEST

If the results of the ADF test show that both spot and futures price series are stationary, then the Granger causality test (Granger, 1969) can be used for determining whether one price series "Granger-causes" the other price series; other examples are for causality and "exogeneity" relations like between (export or financial) development and economic growth (Hsiao, 1987; Calderon and Liu, 2003). The Granger causality test for checking the dependence between two variables, $Y_t$ (i.e., the spot price series) and $X_t$ (i.e. the index futures price series), has the following form

$$Y_t = c + \sum_{i=1}^{n}\theta_i X_{t-i} + \sum_{j=1}^{m}\delta_j Y_{t-j} + \varepsilon_t, \tag{3.3.1}$$

where the null hypothesis can be expressed as $\sum_{i=1}^{m} \theta_i = 0$ or $X_t$ does not cause $Y_t$ (Asteriou and Hall, 2016). Then, the F-statistics value is computed for determining whether the null hypothesis is unacceptable. If the F-statistics value is larger than the F-critical value, the null hypothesis is rejected and one may conclude that $X_t$ causes $Y_t$.

All computations were carried out employing EViews 9.5 (http://www.eviews.com/home.html).

## 4. DATA COLLECTION AND DESCRIPTIVE STATISTICS

All the daily closing spot and futures prices data for the CSI 300 index were collected from the Wind Information website (http://www.wind.com.cn/en/Default.html). Daily spot and futures returns are calculated from

$$R_t = 100 * [\ln(P_t) - \ln(P_{t-1})] \tag{4.1}$$

where $P_t$ denotes the closing spot and futures prices of CSI 300 index on day *t*. Applying daily return series in the empirical analysis of volatility may be beneficial to partially eliminate unit roots and heteroscedasticity in regression models (Engle, 1982). In order to touch upon the 4 aspects measuring the "impact", we have to examine data before and after the CSI 300 stock index futures introduction, but for completeness, the overall time interval as well.

Analyzing CSI 300 spot market volatility, with data starting as early as April 08, 2005, allows for observing the dynamic changes in daily price volatility over the recent fifteen years or so. The lowest values of the spot price volatility occur after the CSI 300 stock index futures launching. Table 4.1 displays the distribution characteristics for the CSI 300 daily returns, over the 2674 first observations; notice that their distribution is leptokurtic and heavy-tailed, i.e., Kurtosis > 3 and Skewness < 0. The CSI 300 daily returns series (not shown) is found to display volatility clustering (Mandelbrot, 1963). The large clustering values (in 2008 and 2015) are known to correspond to financial crashes (Iori, 1999; Shen et al., 2009; Mighri and Mansouri, 2013; Islam, 2014). Thus, the assumption of constant variance is surely not appropriate; moreover due to asymmetric responses, it is surely preferable to employ a TGARCH approach. The results of the analysis are found in Section 5.

Table 4.1 Main descriptive statistical characteristics of CSI300 Index Returns (%) (rounded to four decimals), since its origin, over a long time interval, and more specifically during 3 years before and after the introduction of the Index Futures, on April 16, 2010; N. obs. is the number of observations.

|  | N. obs. | Std.dev | Mean | min. | Max. | Skewness | Kurtosis |
|---|---|---|---|---|---|---|---|
|  |  |  |  |  |  |  |  |
| 08/04/2005-08/04/2016 | 2674 | 1.9032 | -0.3762 | -9.6949 | 8.9310 | -0.5171 | 6.0726 |
| 16/04/2007-16/04/2010 | 734 | 2.4603 | 0.0041 | -8.4560 | 8.9310 | -0.3006 | 4.0783 |
| 16/04/2010-19/04/2013 | 730 | 1.4171 | -0.0401 | -6.4164 | 4.9256 | -0.1304 | 4.6505 |

Next, the daily closing prices of CSI 300 index, from 16/04/2007-19/04/2013, are used to detect the differences of spot price fluctuations before and after launching CSI 300 stock index futures, on 16/04/2010. This amounts to 1463 observations, distributed over a so called Sample I [16/04/2007 - 16/04/2010] and a Sample II [16/04/2010 - 19/04/2013]. The end date of Sample II is April 19, 2013, because the CSI 300 index futures contract delivery date in April 2013 (IF1304) was on 19/04/2013. The daily spot prices have a slight decrease around the onset of index futures trading (i.e. the second quarter of 2010). The price reduction during this period may be attributable to the short-run effects of price transmission and volatility spillovers (Xie and Huang, 2014). The main statistical characteristics are also given in Table 4.1. Histograms of the returns are displayed in Fig. 4.1 and Fig. 4.2

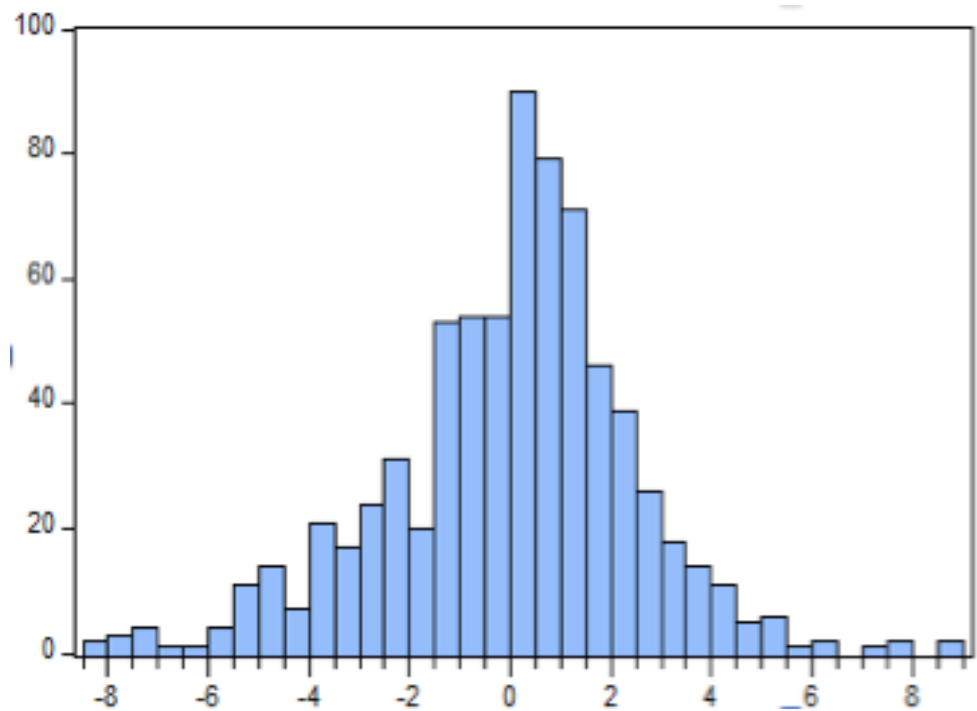

Fig. 4.1 Histogram of the daily returns over 3 years before the launching of the CSI300 FI.

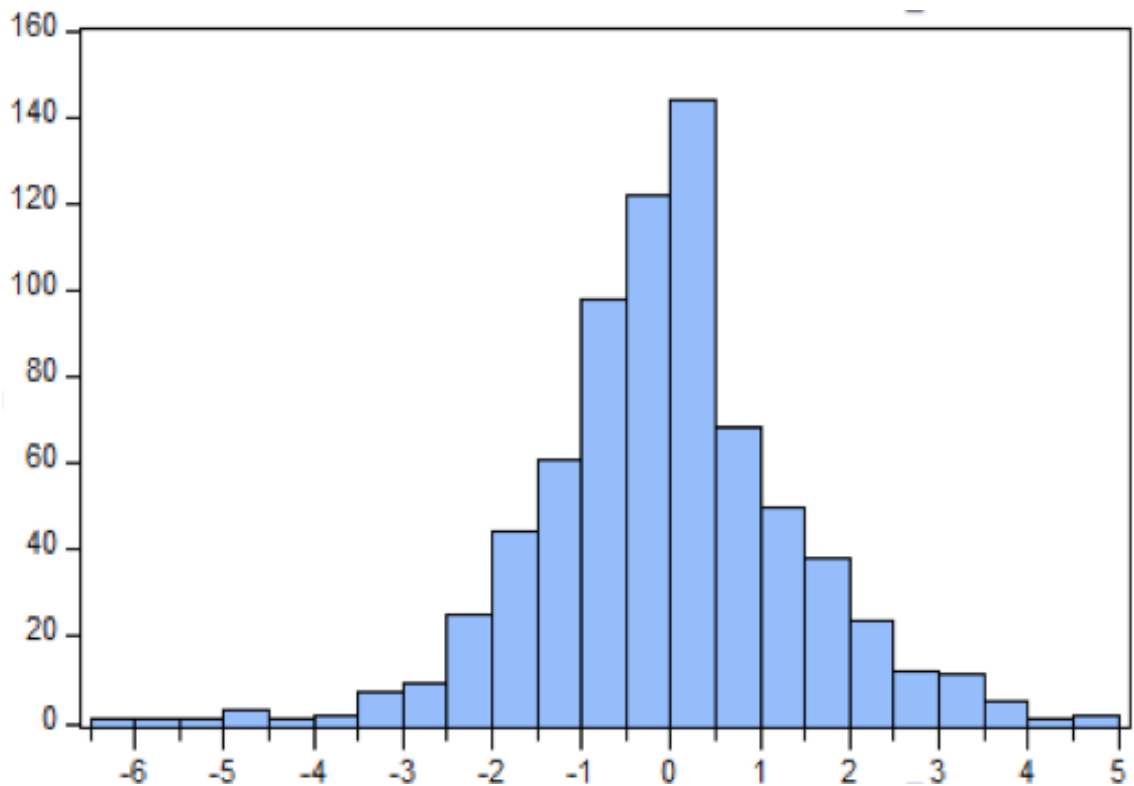

Fig. 4.2 Histogram of the daily returns over 3 years after the launching of the CSI300 FI.

Comparing the return volatility statistical characteristics (in Table 4.1) before and after the introduction of the CSI 300 index futures on April 16, 2010 reveals that the level of return volatility is lowered after issuing index futures contracts: the standard deviation is reduced from 2.4603 to 1.4171. Also, there is a marked spread reduction for the CSI 300 index returns, leading to a more peaky distribution, after the introduction of the CSI 300 IF. The geometric mean, applicable to "measure the average rate of growth" (Watsham and Parramore, 1997, p. 54), decreases from a positive (0.0041) to a negative (-0.0401) value.

The data on Autocorrelations (AC) and Partial Autocorrelations (PAC) of daily returns before and after launching the CSI 300 stock index futures are next discussed. Recall that autocorrelations (AC) can appear in a time series model, when "the error in one period may affect the error in the next time period" (Asteriou and Hall, 2016, p. 157), while partial autocorrelations (PAC) refer to the degree of association between "the current observation of a variable $X_t$" and "successive lagged values of that variable $X_{t-1},…, X_{t-k}$" (Watsham and Parramore, 1997, p. 240). According to the statistical characteristics reported in Table 4.2, there is a slight decrease in the autocorrelation and partial autocorrelation after the introduction of the CSI 300 index futures: indeed, the coefficients of the autocorrelation (AC) and partial autocorrelation (PAC) at lags between 1 and 36, for the 3 years after the introduction of the Index Futures are indeed slightly lower than those corresponding to CSI 300 before April 2010. The Q-statistics does not point to any anomaly, considering the pertinent $\chi^2$.

In summary, at this point, it is observed that there is a high degree of stock price variability before issuing the CSI 300 index futures contract, while the introduction of futures trading appears to bring a beneficial influence to price stabilization in the underlying spot market.

**Table 4.2 Main statistical characteristics of Auto- and Partial Auto-Correlation (AC and PAC) of CSI 300 Index Returns before Index Futures introduction, over 16/04/2007-14/04/2010, after Index Futures introduction, over 16/04/2010-19/04/2013, and over 08/04/2005-08/04/2016, for up to 36 months time lag; thus 35 degrees of freedom (df).**

|  | 16/04/2007-14/04/2010 | | 16/04/2010-19/04/2013 | | 08/04/2005-08/04/2016 | |
|---|---|---|---|---|---|---|
|  | AC | PAC | AC | PAC | AC | PAC |
| Minimum | -0.0720 | -0.0780 | -0.0880 | -0.0970 | -0.0650 | -0.0620 |
| Maximum | 0.0750 | 0.0710 | 0.0750 | 0.0740 | 0.0720 | 0.0700 |
| Mean | 0.0038 | 0.0032 | -0.0021 | -0.0017 | 0.0090 | 0.0071 |
| Std. Dev. | 0.0391 | 0.0380 | 0.0391 | 0.0385 | 0.0291 | 0.0296 |
| Skewness | -0.0224 | -0.1120 | -0.0316 | -0.2560 | -0.0999 | 0.0014 |
| Kurtosis | -1.0690 | -1.0298 | -0.0701 | 0.1804 | 0.0897 | -0.1741 |
| Q-stat | 41.065 | | 40.218 | | 87.312 | |
| Chi2 (5%,df :35) | 49.7658 | | | | | |

Third, we selected daily spot and futures closing prices necessarily from 16/04/2010 to arbitrarily 08/04/2016, thus 6 yeas after the future index introduction, in order to examine (co-integration and) causality between spot and index futures markets: this amounts to 2904 observations in total. However, the findings are not more anomalous that as found here above. Hence, it may be deduced that a linear combination of futures and spot prices is stationary. Yet, to further check the existence of a stationary linear combination of futures and spot prices, co-integration tests are to be performed; see Section 5.3.

5. **REGRESSION ANALYSIS**

5.1. **SPOT MARKET VOLATILITY**

The Augmented Dickey and Fuller test is classically used to detect the presence of unit roots. The ARCH effects can be measured by the Lagrange multiplier (LM) test (Engle, 1982). In this Section 5.1 we report findings with respect to the CSI 300 spot market volatility from April 8, 2005 to April 8, 2016 obtained by estimating the parameters in ARCH and GARCH models, - as if $\gamma = 0$ in equations (3.1.3)-(3.1.4).

According to Table 5.1.1, the Dickey and Fuller (1979) test statistics (i.e. -50.1530) is much smaller than the ADF critical values at 1%, 5% and 10% significant levels. Thus, the null hypothesis of a unit root is rejected; it is concluded that the daily return series is a stationary process.

**Table 5.1.1 Results of the ADF test for the daily returns of CSI 300 Index**

|  |  | t-statistics |
|---|---|---|
| Augmented Dickey and Fuller test |  | -50.1530 |
| Critical values | 1% | -3.9615 |
|  | 5% | -3.4115 |
|  | 10% | -3.1276 |

Moreover, in Table 4.2, one can find that the largest coefficients of autocorrelation (AC) and partial autocorrelation (PAC) are equal to about 0.075 and 0.071 for the "before introduction cases", respectively; browsing through the data such maxima are found to occur at lag 4 (the fourth-order lag). For the "after introduction" series, the largest coefficients of AC and PC are 0.075 and 0.074; they occur at lag 28 and lag 7, respectively. However the next to largest coefficients equal to 0.072 and 0.071 occur at lag 3, respectively. In all cases, the $p$-values ("Prob") at lag 3 and 4 is less than 0.05 for a 95% confidence interval. Thus, it is concluded that the sample of the daily returns has significant autocorrelation and partial autocorrelation at lag 4. It can be deduced that such a finding indicates that the stock price can be predicted on such a time span (Ausloos and Bronlet, 2003; Bronlet and Ausloos, 2003).

The best mean of the return series can be estimated through the LM technique as

$$r_t = 0.0333\, r_{t-4} + \varepsilon_t, \tag{5.1.1}$$

One finds that the variance of a GARCH (1,1) model is

$$\sigma_t^2 = 0.0256 + 0.0554\, \varepsilon_{t-1}^2 + 0.9385\, \sigma_{t-1}^2, \tag{5.1.2}$$

where both present equations (5.1.2) - (5.1.3) are based on equation (3.1.1) and equation (3.1.2). Meanwhile, the sum of the ARCH and the GARCH coefficients equals 0.9939 (i.e. 0.0554+0.9385). The sum of these two parameters is very close to 1, undoubtedly indicating volatility clustering in the daily return series of the CSI 300 index. Selecting 3 (or 4) as the current number of lags leads to conclude that the null hypothesis of no ARCH effect can be rejected, as the probability value ($p$-values) of the F-statistics is found to be less than 0.05, for a 95% confidence interval. Then, doing the ARCH-LM test for the residual series of the GARCH (1,1) model, the output reveals that the $p$-values of the F-statistics is 0.3566 (exceeding 0.05 for a 95% confidence interval), i.e. there is no evidence against the null hypothesis of no ARCH effect. Hence, ARCH effects are eliminated after such a GARCH (1,1) process modeling, - nevertheless a somewhat surprising finding (Jafari et al., 2007).

One may check whether both daily index return series (i.e. before and after the index futures introduction) have unit roots (using ADF test) and ARCH effects (applying an ARCH-LM test). Results (Table 5.2.1) indicate that both return series before and after three years of futures trading introduction are stationary processes, as both ADF test statistics, i.e. -26.380 (before) and -27.437 (after) are much lower than the corresponding ADF critical values at 1%, 5% and 10% significant levels, leading to an evidence against the null hypothesis of a unit root.

Both daily return series have ARCH effects, as both probability values of F-statistics (Table 5.2.2) for the two return series are less than 0.05, for a 95% confidence interval (thus rejecting the null hypothesis of no ARCH effect). Notice that the *p*-value increases to 0.0471 after the index futures introduction, indicating that the null hypothesis of no ARCH effect is more likely to be accepted after issuing the futures contract.

In other words, there is a weaker ARCH effect after CSI 300 stock index futures introduction. Thus, the establishment of the index futures market brings a substantial influence toward reducing volatility clustering in the underlying spot market, - even though ARCH effects are not entirely eliminated.

Table 5.2.1

**Results of ADF Test for the Daily Return Series before and after April 16, 2010**

| TIME PERIOD | ADF Test Statistic | ADF Test Critical Values | | | Probability Value (*Prob*) | Result |
|---|---|---|---|---|---|---|
| | | 1% level | 5% level | 10% level | | |
| Before Three Years (16/04/2007-16/04/2010) | -26.38015 | -3.970621 | -3.415959 | -3.130252 | 0.0000 | Stationary |
| After Three Years (16/04/2010-19/04/2013) | -27.43717 | -3.970691 | -3.415993 | -3.130273 | 0.0000 | Stationary |

Table 5.2.2

**Results of ARCH-LM Test for the Daily Returns Series before and after April 16, 2010**

| TIME PERIOD | F-statistic Value | Probability Value of F-statistic |
|---|---|---|
| Before Three Years (16/04/2007-16/04/2010) | 5.794966 | 0.0006 |
| After Three Years (16/04/2010-19/04/2013) | 1.603828 | 0.0471 |

For completing the argument, a GARCH (1,1) model is studied next on both sides of April 16, 2010. The results are presented in Table 5.2.3. The coefficients $\alpha$ and $\beta$ in Table 5.2.3 represent the volatility clustering and indicate the persistence of shocks due to previous prices of CSI 300 index. From Table 5.2.3, it is deduced that there is a manifestly significant reduction in volatility clustering after introducing the stock index futures, i.e. the coefficient $\alpha$ decreases from a positive 0.05704 (before the index futures introduction) down to a negative -0.021965 (after index futures introduction). Also, the coefficient $\beta$ is reduced from 0.937384 (before) to -0.500689 (after). Therefore, it can be concluded that index futures trading reduces the persistence of shocks due to past prices, and increases the "speed of reaction" toward new prices (i.e., there is a price discovery function effect).

In brief, the CSI 300 index futures introduction is seen to play an important role in spot market stabilization. However, the GARCH (1,1) model does not allow us to observe whether there is an asymmetric impact of good and bad news on the price volatility due to the introduction of the CSI 300 index futures.

Thus, the TGARCH (1,1) model is "finally" attempted in order to investigate asymmetries in positive and negative shocks, along the same lines of thought and tests as the GARCH (1,1) model. The TGARCH (1,1) regression coefficients $\alpha$, $\gamma$, and $\alpha + \gamma$ in Table 5.2.4 represent the effect of "good news", the leverage effect, and the effect of "bad news", respectively. This table reveals that the asymmetries in positive and negative shocks on spot price volatility declined after listing CSI 300 index futures, i.e. the coefficient $\alpha$ goes from 0.003192 to -0.03924, the coefficient $\gamma$, from 0.08833 to 0.07492). Meanwhile, stock index futures trading also results in a weaker leverage effect because the coefficient $\alpha + \gamma$ decreases from 0.091522 (before) to 0.03568 (after).

Chen et al. (2013) attribute the reduction of an asymmetric effect in spot price volatility to price discovery and risk aversion functions in index futures trading. Indeed, the impact of price discovery function results in a faster reaction to new information, thereby decreasing the asymmetries in good and bad news; whilst the risk aversion function in futures trading is helpful to partially eliminate risk, thereby weakening the asymmetric effect on spot market volatility. *A contrario*, Yang et al. (2012) consider that the new stock index futures market does not function well in its price discovery performance.

**Table 5.2.3**

**Regression results of a GARCH (1,1) Model for the Daily Returns**

**Before and After Three Years of CSI 300 Index Futures Introduction on April 16, 2010**

| TIME PERIOD | MODEL | $\alpha$ | $\beta$ |
|---|---|---|---|
| Before Three Years (16/04/2007-16/04/2010) | GARCH (1,1) | 0.057040 ($p$-value: 0.0000) | 0.937384 ($p$-value: 0.0000) |
| After Three Years (16/04/2010-19/04/2013) | GARCH (1,1) | −0.021965 ($p$-value: 0.2536) | -0.500689 ($p$-value: 0.4757) |

**Table 5.2.4**

**Results of TGACRCH (1,1) Model for the Daily Returns**

**Before and After Three Years of CSI 300 Index Futures Introduction**

| TIME PERIOD | MODEL | $\alpha$ | $\gamma$ | $\alpha + \gamma$ |
|---|---|---|---|---|
| Before Three Years (16/04/2007-16/04/2010) | TGARCH (1,1) | 0.003192 | 0.088330 | 0.091522 |
| After Three Years (16/04/2010-19/04/2013) | TGARCH (1,1) | -0.039240 | 0.074920 | 0.035680 |

## 5.3 CO-INTEGRATION AND ERROR-CORRECTION TESTS

This Section 5.3 concerns the results for co-integration and error-correction models (ECM). Indeed, it is important to investigate whether both price and return series are non-stationary employing ADF tests, prior to checking the existence of a long-run equilibrium relationship between index futures and spot markets. The details of the numerical tests are presented in Appendices in order to remain focused on the optimum regression model. Therefore, we only here report the findings in Appendix A1. Both return series (i.e. "index return" and "index futures return") are stationary processes, whereas both logarithmic series of index and index futures prices are non-stationary processes. However, performing ADF tests at the first difference level for these logarithmic series, it is found that they are stationary series.

For completeness, recall that in addition to the two-step procedure of Engle and Granger, the Johansen co-integration test (1991) is another method to measure the co-integration between spot and index futures markets. The numerical results are also reported in Appendix A2.

In addition to co-integration tests, the Error-correlation models (ECM) are useful to detect the short-run dynamics of the relationship in spot and index futures markets. The results, found in Appendix A3, support the finding that spot and futures prices of CSI 300 index are co-integrated.

## 5.4. GRANGER CAUSALITY TEST

The Granger causality test can be used to measure causality between spot and index futures prices (i.e. daily closing prices) of CSI 300 index. Granger (1969) suggests that the Granger causality test is only valid for a stationary time series. Thus, before checking causality in spot and futures markets, it is crucial to ensure that both logarithmic series of daily spot and futures closing prices are stationary processes.

In Section 5.3 it has been shown that both logarithmic series are non-stationary. Yet, both logarithmic series of the first differences are stationary series. According to results presented in Table 5.4.1, the logarithmic series of the daily index futures prices at 1st differences (X) does "cause" the logarithmic series of the daily spot prices at 1st differences (Y), at lags 1 to 10. A complementary discussion is found in Appendix A4, about the p–values for different time lags, pointing to a significant threshold between 2 and 4.

From such a data analysis, it can be therefore concluded that the volatility of CSI 300 index futures prices may be considered to be a "cause" for the fluctuations of the corresponding spot prices at lags 1 and 2, whilst a two-way Granger causality (i.e. the spot price volatility causes futures price volatility and conversely the futures price volatility causes spot price volatility) can be considered to be present at lags between 3 and 10, - namely demonstrating an interesting bi-directional feedback between the two variables. In so doing, a predictability of the short time lag fluctuations is of interest for price prediction (Ausloos and Bronlet, 2003; Bronlet and Ausloos, 2003).

**Table 5.4.1**

**Results of the Granger Causality Tests for CSI300 Index Spot and futures Prices**

| Number of Lag(s) | X does not Granger Cause Y (F-Statistic Value) | X does not Granger Cause Y (p-value) | Y does not Granger Cause X (F-Statistic Value) | Y does not Granger Cause X (p-value) |
| --- | --- | --- | --- | --- |
| 1  | 29.6367 | 6.E-08 | 1.8053  | 0.1793 |
| 2  | 14.2658 | 7.E-07 | 1.5736  | 0.2077 |
| 3  | 10.3810 | 9.E-07 | 13.2804 | 1.E-08 |
| 4  | 7.4889  | 6.E-06 | 10.1959 | 4.E-08 |
| 5  | 5.9679  | 2.E-05 | 7.8617  | 3.E-07 |
| 6  | 5.4964  | 1.E-05 | 7.6757  | 4.E-08 |
| 7  | 4.8773  | 2.E-05 | 7.2320  | 2.E-08 |
| 8  | 4.7492  | 9.E-06 | 6.5623  | 2.E-08 |
| 9  | 4.2764  | 2.E-05 | 5.7953  | 6.E-08 |
| 10 | 3.8325  | 4.E-05 | 5.5911  | 3.E-08 |

**5.5. EMPIRICAL FINDINGS SUMMARY**

Let the above findings be briefly summarized. According to Section 5.1, volatility clustering is persistently present in CSI 300 spot market; meanwhile, there is a significant asymmetric effect, i.e. the asymmetry between positive and negative shocks, on stock price variance. For completeness, we have examined "simple" time series models, but demonstrate that they are not optimal ones.

Yet, results associated with the difference in spot price fluctuations before and after three years of the introduction of CSI 300 index futures trading, employing GARCH (1,1) and TGARCH (1,1) models reveal, in Section 5.2, that volatility clustering in the spot market becomes much weaker following the introduction of CSI 300 stock index futures.

It is shown, as argued, that the TGARCH (1,1) model is the most useful to show the dynamic characteristics of CSI 300 index prices in different periods. Table 5.2.4 illustrates that the asymmetric effect on volatility is reduced after the stock index futures was launched. One may conclude that the introduction of CSI 300 stock index futures brings a beneficial influence to the reduction of the underlying spot market volatility.

Sections 5.3 and 5.4 focus on exploring co-integration and causality in CSI 300 spot and futures markets. Results of Appendices to Section 5.3 illustrate that there is a stationary equilibrium relationship between the two markets. Due to the effect of price discovery function in futures trading, spot market response to new information tends to be faster, and spot prices can be adjusted immediately as the change of spot prices follows the movement of the corresponding index future prices. According to results of the Granger causality test in Section 5.4, one may conclude that CSI 300 index futures market plays a more dominant role than the underlying spot market in determining stock prices, although there is a bi-directional Granger-causality between the two markets.

*In fine*, we do conclude that the introduction of index futures trading in China plays an important role in the spot market stabilization.

## 6. CONCLUSIONS

Our purpose has been to add an original contribution to existing studies by investigating the impact of stock index futures introduction on spot market volatility in three key ways. First, although some existing literature investigates the impact of the CSI 300 index futures on the underlying spot market adopting several ARCH based models, like the Generalized Autoregressive Conditional Heteroscedastic (GARCH) models (Tian and Zheng, 2013; Xie and Huang, 2014; Yang et al., 2012; Zhang, 2015), several others, like the threshold GARCH (TGARCH) model has not been used. However, we have pointed out that the latter model is useful, optimal, for checking whether or not there is an asymmetric response of the spot market volatility upon positive or negative shocks. Secondly, based on some previous studies which focus merely on GARCH family of models (see Bohl, et al., 2015; Wu, 2011), we have also considered not only to test the equilibrium relationship between index futures and spot prices by the co-integration and error correction models (Gosh, 1993; Wahab et al., 1993; Duan and Pliska, 2004; Dempster et al., 2008; Chiu et al., 2015), but also to detect the causality between two prices through the Granger causality test. Thirdly, the time interval chosen for our investigations is much longer than in any other data examined so far: it spans from 8th April 2005 through 8th April 2016, i.e. 11 years in total.

In so doing, beside the empirical findings presented in Section 5, this paper has also provided some literature review in Section 2, indicating the various "conclusions" up to now. We find that our results are different from previous studies (introduced in Section 2). In brief, recall that Aggrawall (1988) and Board et al. (2001) state that index future trading does not change the spot market volatility. In contrast, Furbush (1989), Schwert (1989), and Cheung and Ng (1990) support the view that the introduction of index futures exacerbates the volatility of the spot market (on the US market). Besides, Chang et al (1999) find an increased volatility in the Japanese stock market. In addition, Yang et al. (2012) also report an increased spot market volatility after the CSI 300 index futures was launched in China. Chen and Zhang (2015) find that *"the CSI 300 index futures price does not "Granger cause" the corresponding spot price"*. In contrast, our findings, through the Granger causality test, are inconsistent with the conclusions of Chen and Zhang (2015): we fully underline that there is a bi-directional Granger-causality between futures and spot market in China.

We do conclude that the introduction of index futures trading in China plays an important role in the spot market stabilization. One might appreciate that the finding differs from a classical argument which is the common belief that the more restrictive trading and regulatory constraints, are, like on the Chinese market, the less liquidity and efficient financial markets are, whence the more likely it is that asset prices would deviate from their true values.

We attribute the differences between such conclusions in previous research and our paper to two main factors: (1) our empirical research covers a longer analysis period than previous studies and (2) our study applies a variety of different time series regression models to investigate the volatility linkage between futures and spot markets, - whence we obtain much consistency, due to more sophisticated and complementary techniques.

However, we are aware that there are two main restrictions before claiming universality and indubitable conclusions for our findings. First, the findings so obtained might be only valid for the Chinese stock market. Due to the impact of different situations, like entry requirements and specific government regulations, the results regarding the relationship between spot and futures markets in China might be opposed to the findings about other markets (Chen et al., 2013). No need to say that other developed and emerging markets could be studied within our methodology. Secondly, data collection only concerns the daily closing spot and futures prices of CSI 300 index (also called the low-frequency data). Martens (2002) states that daily prices cannot disclose the dynamic features of financial asset prices in real time, whereas intraday prices, particularly the 5-minute intraday prices, can rapidly and accurately respond to different kinds of information behavior in the stock market. It might be so. Thus, an open question pertains to the fate of intraday prices (instead of daily prices) dynamic changes for related spot and index futures volatility.

We also admit that one cannot insure that our findings are of wholly universal nature. Indeed, the literature review has indicated some discrepancy between authors. We dare to claim that some discrepancy arises from the methodology more than from the data. In order to search for universality, other markets should be investigated, we insist, along a TGARCH model. Besides, we are also aware that effects due to the introduction of a future index in mature markets took place some time ago. The effect of such a time of future index introduction might be relevantly reconsidered. Another relevant point might be of interest, but is hard to measure: the Chinese stock market might be somewhat matured, whence our findings might have been influenced by the globalization having led to a change in the "economic environment". It might be the case in China (Zhang, 2018). We hope that less mature markets might profit from the above findings.

## Acknowledgements

## "later on"

=====================================

## The authors claim to have no conflict of interest

This research did not receive any specific grant from funding agencies in the public, commercial, or not-for-profit sectors.

# APPENDICES

## A1. Appendix to Section 5.3 concerning the results for co-integration tests

To partially remove unit roots and heteroscedasticity in both models, both spot and futures prices series are converted to their corresponding logarithmic series: they are called "$ln(\text{CSI300 index prices})$" and "$ln(\text{CSI300 index futures prices})$".

Prior to checking the existence of a long-run equilibrium relationship between index futures and spot markets, it is important to investigate whether both price and return series are non-stationary employing ADF tests. According to result displayed in Table A.5.3.1, both return series (i.e. "index return" and "index futures return") are stationary processes, whereas both logarithmic series of index and index futures prices are non-stationary processes. Then, doing ADF tests at the first difference ($\Delta$…) level for these two logarithmic series, results displayed in Table A.5.3.1, indicate that the two logarithmic series at the first difference level are stationary series. Futures and spot prices have the same order of integration (i.e. the first-order); next, co-integration tests are carried out in the following analysis utilizing the two-step procedure of Engle and Granger (1987). From Table A.5.3.1, the long-run relationship between the two variables can be expressed as

$$ln(\text{CSI300 Index Prices}) = -0.0509 + 1.0066\, ln(\text{CSI300 Index Futures Prices}) + \varepsilon_t, \quad (A.5.3.1)$$

where Eq. (A.5.3.1) is based on Eq. (3.2.3). The coefficient $\beta$ is positive (i.e. = 1.0066) and is statistically significant at the 5% level (i.e. $p$-value < 0.05). It may be concluded that there is a long-run relationship between spot and futures markets. Subsequently, establishing a residual series for both logarithmic series of spot and futures prices results in finding through the ADF test that the residual series is also a stationary process (Table A.5.3.1), since the ADF test is then found equal to -5.5443, much below the usually called critical values (as those given in Table A.5.3.1).

**Table A.5.3.1**

**ADF Tests for CSI 300 Index and for First Differences (Index and Index Futures) Prices**

|  | ADF | ADF critical value | | | result |
|---|---|---|---|---|---|
|  |  | 1% | 5% | 10% |  |
| ln(Index Prices) | -1.8901 | -3.9644 | -3.4129 | -3.1285 | nonstationary |
| ln(Index Futures Prices) | -2.0713 | -3.9644 | -3.4129 | -3.1285 | nonstationary |
| Index Returns | -36.7427 | -3.9644 | -3.4129 | -3.1285 | stationary |
| Index Futures Returns | -37.3705 | -3.9644 | -3.4129 | -3.1285 | stationary |
|  |  |  |  |  |  |
| $\Delta[ln(\text{Index Prices})]$ | 36.9135 | -3.9644 | -3.4129 | -3.1285 | stationary |
| $\Delta[ln(\text{IndexFuturesPrices})]$ | 37.5144 | -3.9644 | -3.4129 | -3.1285 | stationary |

## A2. Appendix to Section 5.3 on Johansen Co-integration Test

For completeness, recall that in addition to the two-step procedure of Engle and Granger, the Johansen (1991) co-integration test (Maddala, 2001) is another method to measure the co-integration between spot and index futures markets. The Johansen co-integration test consists of Trace and Max-eigenvalue tests. Results indicate that both Trace and Max-eigenvalue tests obey a co-integrating equation at the 0.05 significance level.

## A3. Appendix to Section 5.3 on Error correction models (ECM)

In addition to co-integration tests, the Error correction models (ECM) are useful to detect the short-run dynamics of the relationship in spot and index futures markets. The relationship between two markets can be expressed as $\Delta Y_t = a_0 + b_1 \Delta X_t - \pi \hat{u}_{t-1} + \varepsilon_t$ (i.e. equation 3.2.4). Table A.5.3.6 (I) reveals that $a_0$ (i.e. a constant) and $\Delta \ln(\text{Index Prices})_{t-1}$ are not statistically significant at the 5% level, thereby suggesting that one could-remove these two parameters from the model. The adjusted results of the error correction model are presented in Table 5.3.6(II). According to the output of Tables 5.3.6(II), the error correction model can be written as

$$\Delta \ln(\text{Index Prices})_t = 0.8278 \Delta \ln(\text{Index Futures Prices})_t - 0.2014 \hat{u}_{t-1} + \varepsilon_t, \qquad (5.3.2)$$

where Eq. (5.3.2) is associated with the short-run dynamics relationship between spot and futures markets. Results in Table A.5.3.6 (II) show that $\hat{u}_{t-1}$ (representing the error-correction term) is highly significant at the 5% level, supporting the findings in Section 5.3 that spot and futures prices of CSI 300 index are co-integrated. Note that the error-correction term, $\hat{u}_{t-1}$, is the residual from the co-integrating relationship, lagged 1 time period. The coefficient of $\hat{u}_{t-1}$ is negative, allowing us to conclude that the error correction model is stable. Moreover, this value, -0.2014, suggests a 20.14% movement back towards equilibrium following a shock to the model for a 1 unit time *t* lag.

Table A.5.3.6 (I) and Table A.5.3.6 (II) Results of the Error-correction Model

(I). The Original Results (II). the Adjusted results of the Error-correction Model

| I. | Coeff. | Std. Err. | t-stat |
|---|---|---|---|
| | | | |
| $a_0$ | 8.58 e-06 | 0.0002 | 0.00502 |
| $\hat{u}_{t-1}$ | -0.1993 | 0.0149 | -13.3741 |
| $\Delta[\ln(\text{Index Prices})]$ | 0.0193 | 0.0103 | 1.8773 |
| $\Delta[\ln(\text{Index Futures Prices})]$ | 0.8270 | 0.0093 | 89.3072 |

| II. | Coeff. | Std.Err. | t-stat |
|---|---|---|---|
| | | | |
| $\hat{u}_{t-1}$ | -0.2014 | 0.0148 | -13.5835 |
| $\Delta[\ln(\text{Index Futures Prices})]$ | 0.8278 | 0.0092 | 89.7237 |

**A4. Appendix to Section 5.4 on Granger causality test**

Let us consider the lags between 1 and 10 of both logarithmic series for the "first differences" in view of testing for Granger causality between spot and futures prices. The results displayed in Table 5.4.1 indicate that the logarithmic series of the daily index futures prices at 1st differences (X) does "cause" the logarithmic series of the daily spot prices at 1st differences (Y), at lags 1 to 10, since all the $p$-values (third column of Table 5.4.1) are less than 5% for a 95% confidence interval. In the same Table, one indicates that Y does not "cause X at lags 1 and 2, because the p-values at these two lags (see the fifth column) are larger than 0.05 (whence is not statistically significant); whereas the p-values at lags between 3 and 10 are smaller than 0.05, i.e. rejecting the null hypothesis, since these are therefore statistically significant.